\documentclass{article}
\usepackage[T1]{fontenc}
\usepackage[utf8]{inputenc}
\usepackage{microtype,graphicx,booktabs}
\usepackage[hyperfootnotes=false]{hyperref}

\usepackage[preprint]{icml2026}
\usepackage{amsmath,amssymb,mathtools,amsthm}
\usepackage{xurl}
\usepackage{algorithm,algorithmic}
\newcommand{\bbR}{\mathbb R}
\newcommand{\Exp}{\mathbb E}
\newcommand{\Prob}{\mathbb P}
\newcommand{\indic}[1]{\mathbf 1\{#1\}}
\newcommand{\ip}[2]{\langle #1,#2\rangle}
\newcommand{\norm}[1]{\lVert #1\rVert_2}
\newcommand{\cmax}{c_{\max}}
\DeclareMathOperator{\conv}{conv}

\DeclareMathOperator{\Expl}{Expl}
\newtheorem{theorem}{Theorem}[section]
\newtheorem{lemma}[theorem]{Lemma}
\newtheorem{proposition}[theorem]{Proposition}
\newtheorem{corollary}[theorem]{Corollary}
\theoremstyle{remark}
\newtheorem{remark}[theorem]{Remark}

\usepackage{xcolor}
\definecolor{linkturquoise}{RGB}{0,128,128}
\hypersetup{
    colorlinks=true,
    linkcolor=linkturquoise,
    citecolor=linkturquoise,
    urlcolor=linkturquoise
}

\makeatletter
\renewcommand{\Notice@String}{}
\makeatother
\icmltitlerunning{Last-Iterate Equilibrium Learning in Symmetric Congestion Games under Semi-Bandit Feedback}
\hypersetup{pdftitle={Last-Iterate Equilibrium Learning in Symmetric Congestion Games under Semi-Bandit Feedback},pdfauthor={Katerina Michou, Marina Kontalexi, Dimitris Fotakis},pdfsubject={Research preprint},pdfkeywords={congestion games, semi-bandit feedback, consensus regularization, last-iterate convergence}}
\begin{document}
\onecolumn
\icmltitle{Last-Iterate Equilibrium Learning in Symmetric Congestion Games under Semi-Bandit Feedback}
\begin{icmlauthorlist}
\icmlauthor{Katerina Michou}{ntua}
\icmlauthor{Marina Kontalexi}{ntua,archimedes}
\icmlauthor{Dimitris Fotakis}{ntua,archimedes}

\end{icmlauthorlist}

\icmlaffiliation{ntua}{National Technical University of Athens, Athens, Greece}
\icmlaffiliation{archimedes}{Archimedes/Athena RC, Athens, Greece}

\icmlcorrespondingauthor{Katerina Michou}{katerinmichou@gmail.com}

\vskip 0.2in

\printAffiliationsAndNotice{}
\begin{abstract}
We study last-iterate learning of mixed Nash equilibria in atomic symmetric congestion games under resource semi-bandit feedback, where players observe both the costs and the loads of their selected resources.
For strictly increasing latency functions, the restriction of the potential function to symmetric marginal profiles is strongly convex. However, players' independent stochastic updates need not preserve this symmetry without explicit synchronization.
We show that a quadratic consensus penalty makes the potential globally strongly convex while making the symmetric equilibrium the unique minimizer of the transformed potential.
We propose a variant of online projected gradient descent that achieves $O(t^{-2/3})$ squared distance to equilibrium and $O(t^{-1/3})$ maximum deviation gain in expectation, for every $t\ge1$ and for any initialization. The played profile also converges to equilibrium almost surely. 
\end{abstract}

\section{Introduction}

In an atomic congestion game, each player chooses a feasible set of resources and pays the sum of their load-dependent costs. For example, a driver chooses a route and experiences delays on the roads along it. The same model captures competition for shared resources in communication networks and load balancing. These applications have motivated extensive work on the structure, computation, and efficiency of equilibria~\citep{koutsoupias_worst_1999,gairing_structure_2005,mavronicolas_price_2007,fotakis_structure_2009,christodoulou_price_2005,awerbuch_price_2005,roughgarden_intrinsic_2015}.
Finite exact potential games are precisely the games that are isomorphic to congestion games~\citep{monderer_potential_1996}. Rosenthal's potential guarantees that unweighted congestion games admit a pure Nash equilibrium (PNE) and that every sequence of strict unilateral improvements terminates~\citep{rosenthal_class_1973}. 
Computing a PNE is PLS-complete in general, and the hardness extends to multiplicative approximations~\citep{fabrikant_complexity_2004,skopalik_inapproximability_2008}. Symmetric network congestion games are a notable special case in which a PNE can be computed in polynomial time~\citep{fabrikant_complexity_2004}. 

In repeated settings, where players move independently and receive full counterfactual feedback, convergence to a PNE depends on the update rule and the game.
\citet{kleinberg_multiplicative_2009} show that continuous-time multiplicative updates approach a set of weakly stable equilibria, which are pure for generic congestion costs. \citet{palaiopanos_multiplicative_2017} show that a discrete-time exponential-weights update can cycle or exhibit chaos even in a two-player congestion game. 

Under partial feedback, \citet{heliou_learning_2017} establish almost-sure convergence of Exponential-Weights (EW) to a Nash equilibrium or an approximation thereof in generic potential games. Following this line of work, \citet{giannou_survival_2021} show that, under suitable assumptions on the feedback model and learning parameters, strict Nash equilibria are stochastically asymptotically stable under discrete-time follow-the-regularized-leader (FTRL) dynamics (a superclass of EW), whereas mixed Nash equilibria are not.

These results motivate asking: under which learning dynamics and how quickly can the mixed strategies used by the players approach a (mixed) Nash equilibrium? Under resource semi-bandit feedback, \citet{panageas_semi_2023} give a polynomial bound on exploitability averaged over the learning trajectory. Their result guarantees a small expected deviation gain at a uniformly sampled round. 

\paragraph{Contribution and techniques.}
We study symmetric atomic congestion games with a common feasible strategy set and strictly increasing, possibly nonlinear resource costs. Our target is the symmetric mixed Nash equilibrium, in which all players use the same resource marginals. The intuition is that if all players increase their probability of using a resource, the expected cost of using it rises. This makes the extension of Rosenthal's potential to mixed strategies strongly convex on symmetric profiles, where its first-order optimality condition is exactly the Nash condition. In Proposition~\ref{prop:symmetric-minimizer-mne}, we show that this equilibrium always exists and is unique in marginal space.

However, independent updates need not preserve symmetry, and the strong convexity of the diagonal potential does not extend to the full marginal space. At symmetric profiles, the potential exhibits negative curvature in directions corresponding to disagreement between players' marginals. This motivates adding a quadratic consensus penalty that introduces positive curvature in these directions, while leaving the potential and its gradient unchanged at symmetric profiles. In Theorem~\ref{thm:symmetric-consensus-nonlinear}, we prove that for a sufficiently large penalty, the resulting potential is strongly convex on the full product of marginal polytopes and that its unique minimizer is the symmetric equilibrium. To estimate the correction, a player needs to know the player count $n$, her own marginals, and the realized costs and loads of the resources she selected. The loads reveal how many opponents used each selected resource, so players can estimate the correction without exchanging strategies.

Once all players adopt Algorithm~\ref{alg:consensus}, they independently update their strategies using estimates of the regularized potential's gradient, and we prove that they are guaranteed to converge to the symmetric mixed equilibrium at a rate of $O(t^{-2/3})$ in squared Euclidean distance and $O(t^{-1/3})$ in maximum deviation gain, in expectation, for any iterate $t$ and from arbitrary initialization. The anytime guarantee is established in Theorem~\ref{thm:symmetric-consensus-anytime}, which also proves almost-sure convergence of the played marginal profile. With the covering exploration marginals, the convergence bounds scale polynomially with the number of players and resources, with no dependence on the potentially exponential number of feasible pure strategies. Under their respective assumptions, this provides a last-iterate guarantee, in contrast to the $O(T^{-1/5})$ rate for time-averaged exploitability in~\citep{panageas_semi_2023}. We additionally establish $O(T^{2/3})$ expected realized external regret in self-play (Corollary~\ref{cor:symmetric-self-play-regret}). To our knowledge, this is the first polynomial last-iterate guarantee from arbitrary initialization for learning a possibly mixed Nash equilibrium in this setting under semi-bandit feedback with observed resource loads. It is of independent interest that while restricting the set of desired equilibria imposes a more stringent target, it allows for stronger and faster convergence guarantees due to the structure of the refined target set. The consensus regularization framework and the structural results underlying our analysis originate in the diploma thesis of Michou~\citep{michou_thesis_2026}.

\subsection{Related work}

\paragraph{Learning under congestion feedback.}
\citet{cui_learning_2022} obtain polynomial sample bounds for approximate equilibria using decentralized Frank--Wolfe with exploration, yielding best-iterate guarantees. \citet{panageas_semi_2023} improve the semi-bandit bounds using projected gradient updates on time-varying marginal polytopes and importance-weighted resource-cost estimates. They bound time-averaged exploitability and also establish adversarial no-regret guarantees. \citet{dadi_polynomial_2024} obtain time-averaged exploitability bounds using only the total cost of the selected action. We build on the marginal representation and resource-wise cost estimator of \citet{panageas_semi_2023}, but use observed resource loads to estimate the consensus correction and establish last-iterate convergence. Unlike \citet{panageas_semi_2023}, who perform projected gradient updates on time-varying marginal polytopes, we incorporate exploration through mixing while keeping the projection domain fixed.

\paragraph{Last-iterate learning in potential games.}
\citet{heliou_learning_2017} establish almost-sure convergence to a Nash equilibrium in generic finite potential games under payoff feedback and suitably decreasing exploration.
Under full-information feedback, \citet{anagnostides_last-iterate_2022} show that mirror descent visits an $\varepsilon$-equilibrium within $O(\varepsilon^{-2})$ iterations.
More recently, \citet{anagnostides_convergence_2026} study simultaneous regret matching plus in symmetric potential games. Their symmetry-based result relies on identical initialization and payoff vectors to preserve symmetry throughout learning.

\paragraph{Other approaches to last-iterate convergence.}
\citet{mertikopoulos_learning_2019} establish almost-sure convergence under global variational stability using noisy gradient feedback. Last-iterate guarantees under bandit feedback have also been established for games satisfying strong monotonicity or suitable monotonicity and regularity assumptions~\citep{ba_strong_2025,dong_monotone_2025}. The cost field of a congestion game need not be monotone on the full strategy space; our consensus correction introduces strong monotonicity while preserving the symmetric equilibrium.
\citet{cai_average-iterate_2025} provide a different approach through a reduction from average-iterate to last-iterate convergence, under assumptions including joint affinity in the opponents' strategies.
These assumptions need not hold for general nonlinear congestion costs, and a bound on trajectory-averaged exploitability does not directly yield a bound for the averaged strategy profile for the congestion games setting.

\section{Model and Preliminaries}
\label{sec:model}
\paragraph{Game and marginal representation.}
There are $n\ge2$ atomic, unweighted players and a finite resource set $\mathcal E$ of size $m\ge1$. Each player chooses a pure strategy $p\in\mathcal P\subseteq2^{\mathcal E}$. All players share the same nonempty action set $\mathcal P$. Without loss of generality, we assume that every resource belongs
to at least one feasible action, i.e., $\mathcal E=\bigcup_{p\in\mathcal P}p$. A mixed strategy induces a vector $x_i$ of resource marginals in the polytope
\[
 \mathcal X=\conv\{\mathbf1_p:p\in\mathcal P\}\subseteq[0,1]^m.
\]

By Carathéodory's theorem, every point in $\mathcal X$ can be represented by a distribution supported on at most $m+1$ pure strategies (see also Theorem~3.3 of \cite{panageas_semi_2023}).
 We write $x=(x_1,\ldots,x_n)\in\mathcal X^n$. At each round, players independently sample pure strategies according to their current mixed strategies. Resource-use indicators within a player's sampled strategy may be correlated. For a fixed marginal profile $x$, let $Z_{i,e}$ indicate whether player $i$ uses resource $e$, so $\Prob(Z_{i,e}=1)=x_{i,e}$, and let $L_e=\sum_i Z_{i,e}$.

\paragraph{Latencies and information.}
Each resource has a latency function
$c_e:\{1,\ldots,n\}\to[0,\cmax]$, where
$\cmax<\infty$. Define
\begin{align*}
 \delta_e&=\min_{1\le k<n}[c_e(k+1)-c_e(k)],\\
 \Delta_e&=\max_{1\le k<n}[c_e(k+1)-c_e(k)].
\end{align*}
We assume $0<\delta_e\le\Delta_e$ for every resource. In each round, a player observes the realized costs and loads of her selected resources, but receives no information about unselected resources or her opponents' strategies. Equivalently, selected-resource costs alone suffice if the player knows the corresponding strictly increasing latency tables, allowing her to recover the loads. Players know $n$, the common strategy set, and the parameters required by their updates.

\paragraph{Expected Costs}
For each player $i$, define the conditional resource cost
\[
g_{i,e}(x)=\Exp\!\left[
c_e\!\left(1+\sum_{j\ne i}Z_{j,e}\right)
\right].
\]
This is the expected cost of resource $e$ conditional on player $i$ selecting it. It depends only on the opponents' marginals $x_{-i}$. Player $i$'s expected cost is therefore
\[
c_i(x_i,x_{-i})=\ip{x_i}{g_i(x)}.
\]

\paragraph{Mixed Nash equilibria.}
A mixed-strategy profile $x^\star=(x_i^\star)_{i\in[n]}$ is a mixed Nash equilibrium (MNE) if
\[
C_i(x^\star)\le C_i(x_i,x_{-i}^\star)
\qquad \forall x_i\in\Delta(\mathcal P),\ \forall i,
\]
where $C_i(x)$ denotes player $i$'s expected cost under the mixed-strategy profile $x$. Since expected costs depend only on resource marginals, and every point in $\mathcal X$ is induced by some mixed strategy, this condition is equivalent to
\begin{equation} \ip{g_i(x^\star)}{x_i-x_i^\star}\ge0\qquad\forall x_i\in\mathcal X,\ \forall i\in[n]. \label{eq:symmetric-block-vi} \end{equation}

\paragraph{Symmetry and equilibrium quality.}
Define the mean $m(x)=\frac{1}{n}\sum_i x_i$ and write $\iota(y)=(y,\ldots,y)$, so the symmetric set is $\mathcal S=\{\iota(y):y\in\mathcal X\}$. For a profile $x$, set
\begin{align}
 \operatorname{gap}_i(x)
 &=\max_{y_i\in\mathcal X}\ip{g_i(x)}{x_i-y_i},\nonumber\\
 \Expl(x)&=\max_{i\in[n]}\operatorname{gap}_i(x).
 \label{eq:exploitability}
\end{align}
Thus $\Expl(x)\le\varepsilon$ means that no player can reduce expected cost by more than $\varepsilon$. In particular, $\Expl(x)=0$ characterizes MNE marginals. Our last-iterate guarantees concern the players' mixed strategies, represented by their resource marginals.

\section{Consensus Regularization}
\label{sec:structure}
Following \citet{panageas_semi_2023}, we consider the extension of Rosenthal's potential to mixed strategies. Let $V_e(0)=0$ and $V_e(\ell)=\sum_{r=1}^{\ell}c_e(r)$ for $\ell\ge1$. We define the  potential and its normalized diagonal restriction by
\begin{equation}
\Phi(x)=\sum_{e\in\mathcal E}\Exp[V_e(L_e)],
\qquad
\Psi(y)=\frac1n\Phi(y,\ldots,y).
\label{eq:potential}
\end{equation}

As established in the proof of Theorem~5.1 of \citet{panageas_semi_2023}, the gradient of the potential with respect to each player's marginals coincides with the gradient of her expected cost. Thus, simultaneous gradient updates by players on their individual expected costs amount to a gradient update on the potential. We first show that, when all players use identical marginals, strictly increasing latencies make the potential strongly convex on the symmetric set. Its unique minimizer then corresponds to the unique symmetric mixed Nash equilibrium in marginal space. However, this convexity does not extend to the full strategy space, and independent updates need not preserve symmetry. To address this, we introduce a quadratic consensus penalty that penalizes disagreement between players' marginals. We show that a sufficiently large penalty makes the regularized potential globally strongly convex, while its unique minimizer remains the symmetric equilibrium of the original game.

\begin{proposition}
\label{prop:cost-potential}
The potential $\Phi$ is separable across resources and affine in each marginal coordinate. Moreover,
\[
\nabla_i\Phi(x)=g_i(x),
\qquad
\nabla\Psi(y)=g_i(y,\ldots,y)
\]
for every player $i$.
\end{proposition}

\begin{proof}
Let $L_{-i,e}=\sum_{j\ne i}Z_{j,e}$. By independence across players,
\[
\Exp[V_e(L_e)]
=x_{i,e}\Exp[V_e(1+L_{-i,e})]
+(1-x_{i,e})\Exp[V_e(L_{-i,e})].
\]
Differentiating with respect to $x_{i,e}$ gives
\[
\frac{\partial\Phi}{\partial x_{i,e}}(x)
=\Exp[V_e(1+L_{-i,e})-V_e(L_{-i,e})]
=g_{i,e}(x).
\]
The identity for $\Psi$ follows by the chain rule and player symmetry.
\end{proof}
\begin{proposition}
\label{prop:symmetric-minimizer-mne}

The diagonal potential $\Psi$ is $(n-1)\min_e\delta_e$-strongly convex on $\mathcal X$ and has a unique minimizer $y^\star$. It holds that a strategy profile $(y,\ldots,y)$ is an MNE profile if and only if $y=y^\star$.
\end{proposition}

\begin{proof}
For a resource $e$ at a symmetric profile $(y,\ldots,y)$, the load is binomial with parameters $n$ and $p=y_e$. Its contribution to $\phi(y)=\Phi(y,\ldots,y)=n\Psi(y)$ is
\[
 \phi_e(p)=\sum_{\ell=0}^n V_e(\ell)\binom n\ell p^\ell(1-p)^{n-\ell}.
\]
Let $b_{k,r}(p)=\binom rk p^k(1-p)^{r-k}$, with out-of-range terms set to zero. The identity $b'_{\ell,n}=n(b_{\ell-1,n-1}-b_{\ell,n-1})$, followed by shifting the finite summation index, gives
\begin{align*}
 \phi_e'(p)
 &=n\sum_{k=0}^{n-1}[V_e(k+1)-V_e(k)]b_{k,n-1}(p)\\
 &=n\sum_{k=0}^{n-1}c_e(k+1)b_{k,n-1}(p),\\
 \phi_e''(p)
 &=n(n-1)\sum_{k=0}^{n-2}[c_e(k+2)-c_e(k+1)]b_{k,n-2}(p).
\end{align*}

Since $\Psi=\phi/n$, we obtain
\begin{equation}
 (n-1)\delta_e
 \le
 \frac{\partial^2\Psi(y)}{\partial y_e^2}
 =(n-1)\Exp[
 c_e(K_e+2)-c_e(K_e+1)]
 \le(n-1)\Delta_e,
 \label{eq:diagonal-curvature}
\end{equation}
where $K_e\sim\operatorname{Bin}(n-2,y_e)$.
As $\Psi$ is resource-separable, its Hessian
is diagonal. Hence $\Psi$ is
$(n-1)\min_e\delta_e$-strongly convex on
$\mathcal X$ and admits a unique minimizer.

By Proposition~\ref{prop:cost-potential}, $\nabla\Psi(y)=g_i(y,\ldots,y)$ for every $i$. Thus, the first-order optimality condition for minimizing $\Psi$ over $\mathcal X$ coincides with the MNE condition~\eqref{eq:symmetric-block-vi} at $(y,\ldots,y)$.

\end{proof}
At symmetric profiles, each resource Hessian block of the potential has zero diagonal and identical, strictly positive off-diagonal entries. Hence, it exhibits negative curvature on $\mathbf{1}^{\perp}$, corresponding to disagreement between players' marginals. This motivates the introduction of a quadratic consensus penalty. We define the consensus penalty and regularized objective by
\begin{equation}
R_\rho(x)=\frac{\rho}{2}\sum_i\|x_i-m(x)\|_2^2,\qquad F_\rho=\Phi+R_\rho.
\end{equation}
The penalty has gradient $\nabla_i R_\rho(x)=\rho(x_i-m(x))$ and adds positive curvature in disagreement directions. Both the penalty and its gradient vanish on the symmetric set $S$. We next show that a sufficiently large $\rho$ makes $F_\rho$ strongly convex on the full strategy space.

\begin{theorem}
\label{thm:symmetric-consensus-nonlinear}
Suppose
\begin{equation}
 \rho>\max_{e\in\mathcal E}\left\{\delta_e+\frac{n}{2}(\Delta_e-\delta_e)\right\}.
 \label{eq:threshold}
\end{equation}
Then $F_\rho$ is $\lambda_\rho$-strongly convex on $\mathcal X^n$, where
\begin{equation}
 \lambda_\rho=\min_{e\in\mathcal E}\min\left\{(n-1)\delta_e,\rho-\delta_e-\frac{n}{2}(\Delta_e-\delta_e)\right\}>0.
 \label{eq:nonlinear-modulus}
\end{equation}
Its unique minimizer is $x^\star=\iota(y^\star)$, the unique symmetric MNE marginal profile of the original game.
\end{theorem}
\begin{proof}
Fix a resource $e$. For $i\ne j$, differentiating its conditional cost gives
\begin{align*}
 a_{ij}^{(e)}(x)
 &:=\frac{\partial^2\Phi}{\partial x_{i,e}\partial x_{j,e}}(x)\\
 &=\Exp[c_e(2+S_{ij,e})-c_e(1+S_{ij,e})],
\end{align*}
where $S_{ij,e}=\sum_{k\ne i,j}Z_{k,e}$. Since $0\le S_{ij,e}\le n-2$, the increment bounds give $\delta_e\le a_{ij}^{(e)}(x)\le\Delta_e$. The resource Hessian block $H_e(x)$ has these off-diagonal entries and zero diagonal, because the potential is separately affine in each $x_{i,e}$.

Let $J=\mathbf1\mathbf1^\top/n$ and $Q=I_n-J$. For $v\in\bbR^n$, write $s_+=\sum_i\max\{v_i,0\}$ and $s_-=\sum_i\max\{-v_i,0\}$. Start each off-diagonal entry at $\delta_e$. Increasing it can decrease the quadratic form only for a pair of opposite signs, so
\[
 v^\top H_e(x)v\ge\delta_e\bigl[(\mathbf1^\top v)^2-\norm{v}^2\bigr]-2(\Delta_e-\delta_e)s_+s_-.
\]
Since $4s_+s_-=\|v\|_1^2-(\mathbf1^\top v)^2\le n\norm{Qv}^2$, adding the penalty Hessian $\rho Q$ gives
\[
 v^\top(H_e(x)+\rho Q)v\ge(n-1)\delta_e\norm{Jv}^2+\left(\rho-\delta_e-\frac{n}{2}(\Delta_e-\delta_e)\right)\norm{Qv}^2.
\]
The full Hessian is block diagonal across resources, proving~\eqref{eq:nonlinear-modulus}.

Compactness gives a minimizer, and strong convexity makes it unique. Both the objective and $\mathcal X^n$ are invariant under every permutation $P$ of player blocks. Thus $Px^\star$ is also a minimizer, forcing $Px^\star=x^\star$ for every $P$. Hence $x^\star=(y,\ldots,y)$ for some $y\in\mathcal X$. On this set,
\[
 F_\rho(y,\ldots,y)=\Phi(y,\ldots,y)=n\Psi(y).
\]
Its common block must minimize $\Psi$. Proposition~\ref{prop:symmetric-minimizer-mne} identifies it as $y^\star$ and establishes the equilibrium claim.
\end{proof}

\paragraph{Affine special case.}
If $c_e(\ell)=\alpha_e+\beta_e\ell$ with $\beta_e>0$, then $\delta_e=\Delta_e=\beta_e$. The threshold reduces to $\rho>\max_e\beta_e$, and
\[
 \lambda_\rho=\min\{(n-1)\beta_{\min},\rho-\beta_{\max}\},
\]
where $\beta_{\min}=\min_e\beta_e$ and $\beta_{\max}=\max_e\beta_e$. In this case, the resource Hessian is $\beta_e(\mathbf{1}\mathbf{1}^{\top}-I_n)$ at every profile, and the consensus penalty shifts its eigenvalue on $\mathbf{1}^{\perp}$ from $-\beta_e$ to $\rho-\beta_e$.

\section{Semi-Bandit Learning and Convergence}
\label{sec:algorithm}
The previous section establishes that learning the symmetric equilibrium can be formulated as minimizing the strongly convex objective $F_\rho$. The difficulty under semi-bandit feedback is that the gradient of the consensus penalty depends on the marginals of all players, which are not directly observable. We show that the costs and loads of the selected resources suffice to construct an unbiased estimator of the regularized gradient without communication between players. Using this estimator, we develop a projected gradient algorithm with vanishing exploration and establish last-iterate convergence to the symmetric equilibrium from arbitrary initial strategies. We also study the external regret of the resulting dynamics, both in self-play and against opponents who do not follow the algorithm.
\subsection{Gradient Estimation and Learning Algorithm}

To construct importance-weighted gradient estimators and control their second moments, we need a positive lower bound on the probability of selecting each resource. We achieve this by mixing each player's marginal with a fixed exploration marginal that has full support over the resources. The exploration weight decreases to zero over time. Each player updates her marginal by projected gradient descent on $\mathcal X$ and incorporates exploration only when sampling her strategy.

Let $D_{\mathcal X}=\operatorname{diam}(\mathcal X)$ and $D=\sqrt{n}D_{\mathcal X}$ denote the Euclidean diameters of $\mathcal X$ and $\mathcal X^n$, respectively. For each resource $e$, choose a strategy $p_e$ containing it. Then $m^{-1}\sum_e\mathbf1_{p_e}$ is a feasible marginal with every coordinate at least $1/m$. Each player fixes a marginal $q_i\in\mathcal X$ with $q_{i,e}>0$ for every resource $e$, and we write $\bar\mu=\min_{i,e}q_{i,e}$. If each player uses the preceding construction, then $\bar\mu\ge1/m$, even if the chosen covering strategies differ across players.

In round $t$, player $i$ has a hidden marginal $y_i^t\in\mathcal X$ and plays the marginal
\begin{equation}
 x_i^t=(1-\gamma_t)y_i^t+\gamma_t q_i,\qquad 0<\gamma_t\le1.
\label{eq:exploration-mixing}
\end{equation}
Thus $x_{i,e}^t\ge\mu_t:=\bar\mu\gamma_t$ and $\norm{x^t-y^t}\le D\gamma_t$. Write $\Pi_K$ for Euclidean projection onto $K$. Let $\mathcal F_t$ contain the history through round $t-1$ and the marginals before sampling in round $t$. Conditional on $\mathcal F_t$, players independently sample actions with marginals $x_i^t$ and play them simultaneously. Write $Z_{i,e}^t=\indic{e\in p_i^t}$ and $L_e^t=\sum_i Z_{i,e}^t$. Define
\begin{align}
 \widehat c_{i,e}^t
 &=\frac{Z_{i,e}^t c_e(L_e^t)}{x_{i,e}^t},
 \label{eq:cost-estimator}\\
 \widehat m_{i,e}^t
 &=\frac1n\left[x_{i,e}^t+
 \frac{Z_{i,e}^t(L_e^t-1)}{x_{i,e}^t}\right],
 \label{eq:mean-estimator}\\
 \widehat G_{i,e}^t
 &=\widehat c_{i,e}^t+\rho(x_{i,e}^t-\widehat m_{i,e}^t).
 \label{eq:gradient-estimator}
\end{align}
When $Z_{i,e}^t=0$, the observation-dependent numerators in~\eqref{eq:cost-estimator}--\eqref{eq:mean-estimator} are defined as zero and no latency or load on $e$ is queried. 
When $Z_{i,e}^t=1$, the player uses the observed load $L_e^t$. Thus, these expressions require only permitted observations and the player's own marginal.

\begin{lemma}[Consensus estimator]
\label{lem:symmetric-consensus-estimator}
For $0<\mu_t\le\min\{1,\bar\mu\}$, under the conditional sampling rule above, the following hold:
\begin{enumerate}
    \item[(i)] $\Exp[\widehat m_{i,e}^t\mid\mathcal F_t]=m_e(x^t)$ for every $i,e$.
    \item[(ii)] $\Exp[\widehat G_i^t\mid\mathcal F_t]=\nabla_iF_\rho(x^t)$ for every $i$.
    \item[(iii)] $\Exp[\norm{\widehat G^t}^2\mid\mathcal F_t]\le C_\rho/\mu_t$,
\end{enumerate}
where $\widehat G^t=(\widehat G_1^t,\ldots,\widehat G_n^t)$ and
\begin{equation}
C_\rho=nm(2\cmax^2+8\rho^2).
\label{eq:variance-constant}
\end{equation}
\end{lemma}
\begin{proof}
All expectations below condition on $\mathcal F_t$. The identity
\[
 Z_{i,e}^t(L_e^t-1)=\sum_{j\ne i}Z_{i,e}^tZ_{j,e}^t
\]
and independence across players imply
\[
 \Exp[\widehat m_{i,e}^t\mid\mathcal F_t]
 =\frac1n\left[x_{i,e}^t+\sum_{j\ne i}x_{j,e}^t\right]=m_e(x^t).
\]
Similarly, conditioning on player $i$ selecting $e$ leaves the other players' resource indicators unchanged in distribution, giving $\Exp[\widehat c_{i,e}^t\mid\mathcal F_t]=g_{i,e}(x^t)$. Together with $\nabla_iR_\rho(x)=\rho(x_i-m(x))$, this establishes (ii).

For the second moment, bounded latencies give
\[
 \Exp[(\widehat c_{i,e}^t)^2\mid\mathcal F_t]
 \le\frac{\cmax^2}{x_{i,e}^t}\le\frac{\cmax^2}{\mu_t}.
\]
Write $B=Z_{i,e}^t(L_e^t-1)/(n x_{i,e}^t)$. Since $0\le L_e^t-1\le n-1$ whenever $Z_{i,e}^t=1$,
\[
 \Exp[B^2\mid\mathcal F_t]
 \le\frac{(n-1)^2}{n^2x_{i,e}^t}\le\frac1{\mu_t}.
\]
Now $x_{i,e}^t-\widehat m_{i,e}^t=(1-1/n)x_{i,e}^t-B$, so $(a+b)^2\le2a^2+2b^2$, $x_{i,e}^t\le1$, and $\mu_t\le1$ yield
\[
 \Exp[(x_{i,e}^t-\widehat m_{i,e}^t)^2\mid\mathcal F_t]
 \le2+2/\mu_t\le4/\mu_t.
\]
Applying the same scalar inequality to~\eqref{eq:gradient-estimator} gives
\[
 \Exp[(\widehat G_{i,e}^t)^2\mid\mathcal F_t]
 \le(2\cmax^2+8\rho^2)/\mu_t.
\]
Summing over all $nm$ coordinates establishes (iii).
\end{proof}

Importance weighting thus gives an unbiased estimate of the opponents' marginal sum without observing it directly.

\begin{algorithm}[t]
\caption{Consensus-regularized semi-bandit learning }
\label{alg:consensus}
\begin{algorithmic}[1]
\renewcommand{\algorithmicrequire}{\textbf{Input:}}
\REQUIRE Number of players $n$, regularization parameter $\rho$ satisfying~\eqref{eq:threshold}, schedules $(\gamma_t,\eta_t)$, exploration marginal $q_i\in\mathcal X$ satisfying $q_{i,e}>0$ for all $e\in\mathcal E$, and initial marginal $y_i^1\in\mathcal X$.
\FOR{$t=1,2,\ldots$}
\STATE Set $x_i^t=(1-\gamma_t)y_i^t+\gamma_t q_i$.
\STATE Obtain a distribution $\pi_i^t$ over $\mathcal P$ representing $x_i^t$.
\STATE Sample $p_i^t\sim\pi_i^t$ and play simultaneously with the other players.
\STATE Observe the realized cost $c_e(L_e^t)$ and load $L_e^t$ for each $e\in p_i^t$.
\STATE Compute $\widehat G_i^t$ using~\eqref{eq:cost-estimator}--\eqref{eq:gradient-estimator}.
\STATE Update
\[
 y_i^{t+1}=\Pi_{\mathcal X}(y_i^t-\eta_t\widehat G_i^t).
\]
\ENDFOR
\end{algorithmic}
\end{algorithm}

We now present Algorithm~\ref{alg:consensus}, which combines projected gradient updates using the consensus-regularized gradient estimator with vanishing exploration. Its implementation requires representing marginals as distributions over $\mathcal P$ and computing projections onto $\mathcal X$. By Carath\'eodory's theorem, each marginal admits a representing distribution supported on at most $m+1$ pure strategies. The complexity of both operations depends on the representation of $\mathcal P$. 

\subsection{Last-Iterate Convergence}
\label{sec:convergence}

Fix $\rho$ satisfying~\eqref{eq:threshold} and write $x^\star=\iota(y^\star)$. Throughout the analysis, $\lambda_\rho$ denotes the strong-convexity modulus in~\eqref{eq:nonlinear-modulus} and any known positive lower bound may be used in its place.  By the resource Hessian formula in Theorem~\ref{thm:symmetric-consensus-nonlinear}, every absolute row sum is at most $(n-1)\max_e\Delta_e$ for $\nabla^2\Phi$. Since the matrix is symmetric and the penalty Hessian has operator norm $\rho$, we may use the gradient Lipschitz constants
\begin{equation}
 L_\Phi=(n-1)\max_e\Delta_e,\qquad L_F=L_\Phi+\rho
 \label{eq:geometry-constants}
\end{equation}
for $\Phi$ and $F_\rho$, respectively.

\paragraph{Conversion from distance to exploitability.}
The Nash variational inequality at $x^\star$ and
Lipschitz continuity of $\nabla\Phi$ imply that,
for every $x\in\mathcal X^n$,
\begin{equation}
    \operatorname{Expl}(x)
    \le C_{\mathrm{expl}}\|x-x^\star\|_2,
    \qquad
    C_{\mathrm{expl}}
    = L_\Phi D_{\mathcal X}+\sqrt{m}\,c_{\max}.\label{eq:symmetric-exploitability-lipschitz}
\end{equation}
Consequently, by Cauchy--Schwarz,
\[
    \mathbb E[\operatorname{Expl}(x^t)]
    \le C_{\mathrm{expl}}
    \sqrt{\mathbb E\|x^t-x^\star\|_2^2}.
\]
The proof is given in Appendix~\ref{app:expl}.

\begin{theorem}[Anytime last-iterate convergence]
\label{thm:symmetric-consensus-anytime}
Let $0<\gamma_0\le1$. Suppose all players run Algorithm~\ref{alg:consensus} from arbitrary $y_i^1\in\mathcal X$ with
\[
 \gamma_t=\gamma_0t^{-1/3},\qquad \eta_t=\frac{2}{\lambda_\rho t}.
\]
There is a constant $C_{\rm any}$, independent of $t$, such that for every $t\ge1$ the played profile satisfies
\begin{align}
 \Exp[\norm{x^t-x^\star}^2]&\le C_{\rm any}t^{-2/3},\label{eq:anytime-bound}\\
 \Exp[\Expl(x^t)]&\le C_{\rm expl}\sqrt{C_{\rm any}}\,t^{-1/3}.\nonumber
\end{align}
Moreover, $y^t\to x^\star$ and $x^t\to x^\star$ almost surely.
\end{theorem}

The exploration weight $\gamma_t$ affects the convergence analysis in two ways. A larger $\gamma_t$ decreases the second-moment bound of the gradient estimator, but increases the discrepancy between the played marginal $x^t$ and the hidden marginal $y^t$. These two effects contribute terms of order $\eta_t^2/\gamma_t$ and $\eta_t\gamma_t^2$, respectively, to the squared-distance recursion~\eqref{eq:anytime-recursion}. Taking $\eta_t$ proportional to $t^{-1}$ and $\gamma_t$ proportional to $t^{-b}$, these terms have orders $t^{-(2-b)}$ and $t^{-(1+2b)}$. Balancing the two exponents gives $2-b=1+2b$, hence $b=1/3$. Appendix~\ref{app:anytime} gives the proof and the expression for $C_{\rm any}$. Since $D^2\le nm$ and the exploration construction of Section~\ref{sec:algorithm} yields $\bar\mu\ge 1/m$, the convergence bounds depend polynomially on $n$, $m$, $c_{\max}$, $\max_e\Delta_e$, $\rho$, $\lambda_\rho^{-1}$, and $\gamma_0^{-1}$, independently of $|\mathcal P|$.

The rates above are tight for the chosen exploration schedule. Consider two players choosing between parallel links with latencies $c_A(\ell)=\ell$ and $c_B(\ell)=3+\ell$, and let $q_i=(1/2,1/2)$. Link $B$ always costs at least two units more than $A$, so at equilibrium both players choose $A$. However, the exploration mechanism ensures that $x_{i,B}^t\ge\gamma_t/2$, regardless of the hidden iterate $y_i^t$. Consequently, $\Expl(x^t)\ge\gamma_t$ and $\norm{x^t-x^\star}^2=2\sum_i(x_{i,B}^t)^2\ge\gamma_t^2$. Thus, for the exploration schedule $\gamma_t=\gamma_0t^{-1/3}$, both rates in~\eqref{eq:anytime-bound} are attained.

\subsection{External Regret}

Let $Z_i^t=\mathbf1_{p_i^t}$ and $d_{i,e}^t=c_e(1+\sum_{j\ne i}Z_{j,e}^t)$, where $d_{i,e}^t$ is the cost player $i$ would incur by selecting resource $e$ against the opponents' realized actions at round $t$. The realized external regret of player $i$ with respect to the best fixed pure strategy in hindsight is
\[
R_i^{\rm real}(T)=\sum_{t=1}^T\ip{Z_i^t}{d_i^t}-\min_{p\in\mathcal P}\sum_{t=1}^T\ip{\mathbf 1_p}{d_i^t}.
\]
Since the cumulative cost is linear in the comparator and $\mathcal X=\conv\{\mathbf 1_p:p\in\mathcal P\}$, we can equivalently write
\[
R_i^{\rm real}(T)=\sum_{t=1}^T\ip{Z_i^t}{d_i^t}-\min_{u\in\mathcal X}\sum_{t=1}^T\ip{u}{d_i^t}.
\]

\begin{corollary}
\label{cor:symmetric-self-play-regret}
Suppose all players follow Algorithm~\ref{alg:consensus} with the exploration and stepsize schedules of Theorem~\ref{thm:symmetric-consensus-anytime}. Then, for every player $i$,
\[
\Exp[|R_i^{\rm real}(T)|]=O(T^{2/3}).
\]
\end{corollary}
The proof is given in Appendix~\ref{app:regret}.

\begin{proposition}[Failure of adversarial no regret]
\label{prop:symmetric-adversarial-counterexample}
For every $\rho>1$, in the two-player, two-link game $c_A(\ell)=c_B(\ell)=\ell$, let player $2$ always choose $A$. Suppose player $1$ runs Algorithm~\ref{alg:consensus} with a strictly positive exploration marginal, $0<\gamma_t\le1$, and positive stepsizes, with deterministic schedules satisfying
\[
 \gamma_t\to0,\qquad \eta_t\to0,\qquad \sum_t\eta_t=\infty,\qquad \sum_t\eta_t^2/\gamma_t<\infty.
\]
Then
\[
 \frac{\Exp[R_1^{\rm real}(T)]}{T}\longrightarrow 1-\frac1\rho>0.
\]
In particular, the anytime schedule has linear expected external regret against this fixed opponent.
\end{proposition}

\medskip

\begin{remark}
While consensus regularization enables convergence to the symmetric equilibrium in self-play, against arbitrary opponents it may encourage a player to adopt strategies that incur higher costs, as illustrated by Proposition~\ref{prop:symmetric-adversarial-counterexample}. A natural way to address this limitation is to monitor the player's regret and abandon the consensus term when the regret becomes too large. Following the regret-monitoring idea of \citet{cai_average-iterate_2025},
player $i$ tracks the quantity
\[
 \widehat R_i(k)
 =\sum_{t=1}^k C_i^t
 -\min_{u\in\mathcal X}\left\langle u,\sum_{t=1}^k\widehat c_i^t\right\rangle,
\]
where $C_i^t$ is her observed total cost. Motivated by the $O(T^{2/3})$ expected regret bound we established under self play, we can introduce a switching rule: once $\widehat R_i(k)$ exceeds a threshold of order $k^{2/3}$, the player permanently switches to projected gradient descent using only the cost estimator $\widehat c_i^t$, with exploration weights and stepsizes proportional to $s^{-1/3}$ and $s^{-2/3}$, respectively, where $s$ counts the rounds since switching. 
\end{remark}

\section{Discussion and Limitations}
\label{sec:discussion}

Our results show that polynomial last-iterate convergence guarantees can be obtained even when the potential of the original game is nonconvex. Rather than imposing additional convexity or monotonicity assumptions on the game, our approach exploits the structure of the desired equilibrium to introduce these properties through regularization. In particular, we obtain a strongly convex objective without requiring prior knowledge of its minimizer. Despite the limited feedback available to the players and the fact that their independent updates need not preserve symmetry, we obtain both explicit convergence rates for the strategies played at each round and almost-sure convergence to the unique symmetric equilibrium.

Our analysis assumes that players observe the realized loads on the resources they use, know the number of players, and have sufficient information to choose the regularization strength and learning rates. Finally, the regret guarantees apply to self-play; without additional safeguards, the regularized dynamics may incur linear regret against arbitrary opponents.

\section*{Acknowledgments}
The consensus regularization approach developed in this paper originates in  Katerina Michou's diploma thesis \citep{michou_thesis_2026}, completed at the National Technical University of Athens. 
\bibliography{references}
\bibliographystyle{icml2026}
\clearpage
\onecolumn
\appendix
\section{From Distance to Exploitability}
\label{app:expl}
\begin{proof}[Proof of~\eqref{eq:symmetric-exploitability-lipschitz}] For
every player \(i\), let \(y_i(x)\in\mathcal X\) be a best response to
\(x_{-i}\). Then
\[
\begin{aligned}
\operatorname{gap}_i(x)
&=
\ip{\nabla_i\Phi(x)}{x_i-y_i(x)}\\
&=
\ip{
\nabla_i\Phi(x)-\nabla_i\Phi(x^\star)
}{
x_i-y_i(x)
}
+
\ip{
\nabla_i\Phi(x^\star)
}{
x_i-y_i(x)
}.
\end{aligned}
\]
Insert and subtract \(x_i^\star\) in the second inner product. The Nash
VI at \(x^\star\) implies
\[
\ip{
\nabla_i\Phi(x^\star)
}{
x_i^\star-y_i(x)
}
\le0.
\]
Hence
\[
\begin{aligned}
\operatorname{gap}_i(x)
&\le
L_\Phi
\norm{x-x^\star}
\norm{x_i-y_i(x)}
+
\norm{\nabla_i\Phi(x^\star)}
\norm{x_i-x_i^\star}\\
&\le
\left(
L_\Phi D_{\mathcal X}
+
\sqrt m\,\cmax
\right)
\norm{x-x^\star}.
\end{aligned}
\]
Therefore
\begin{equation}
\Expl(x)
\le
C_{\mathrm{expl}}\norm{x-x^\star},
\qquad
C_{\mathrm{expl}}
=
L_\Phi D_{\mathcal X}
+
\sqrt m\,\cmax.
\end{equation}
Taking expectations and applying Cauchy--Schwarz gives the claimed bound in expectation.
\end{proof}

\section{Proof of Last-Iterate Convergence}
\label{app:anytime}
\begin{proof}[Proof of Theorem~\ref{thm:symmetric-consensus-anytime}]
Let $v_t=\norm{y^t-x^\star}^2$. Since $x^\star$ minimizes $F_\rho$ over $\mathcal X^n$, its variational inequality and strong convexity give
\[
 \ip{\nabla F_\rho(y^t)}{y^t-x^\star}\ge\ip{\nabla F_\rho(x^\star)}{y^t-x^\star}+\lambda_\rho v_t\ge\lambda_\rho v_t.
\]
Projection onto $\mathcal X^n$ is nonexpansive. Expand the squared distance before projection and apply Lemma~\ref{lem:symmetric-consensus-estimator} at the played profile to obtain
\[
 \Exp[v_{t+1}\mid\mathcal F_t]\le v_t-2\eta_t\ip{\nabla F_\rho(x^t)}{y^t-x^\star}+\frac{C_\rho\eta_t^2}{\bar\mu\gamma_t}.
\]
By~\eqref{eq:exploration-mixing}, $\norm{\nabla F_\rho(x^t)-\nabla F_\rho(y^t)}\le L_FD\gamma_t$. Using $2L_FD\gamma_t\sqrt{v_t}\le\lambda_\rho v_t+(L_FD\gamma_t)^2/\lambda_\rho$ gives
\begin{equation}
 \Exp[v_{t+1}\mid\mathcal F_t]\le(1-\lambda_\rho\eta_t)v_t+\frac{L_F^2D^2}{\lambda_\rho}\eta_t\gamma_t^2+\frac{C_\rho\eta_t^2}{\bar\mu\gamma_t}.
 \label{eq:anytime-recursion}
\end{equation}

Write $a_t=\Exp[v_t]$. Substituting the schedules gives
\[
 a_{t+1}\le(1-2/t)a_t+Ht^{-5/3},\qquad H=\frac{2L_F^2D^2\gamma_0^2}{\lambda_\rho^2}+\frac{4C_\rho}{\lambda_\rho^2\bar\mu\gamma_0}.
\]
Take $M=\max\{2^{2/3}D^2,3H/4\}$. Since $a_1,a_2\le D^2$, we have $a_t\le Mt^{-2/3}$ for $t=1,2$. For $t\ge2$, assuming $a_t\le Mt^{-2/3}$, we obtain
\[
 a_{t+1}\le Mt^{-2/3}-\tfrac23 Mt^{-5/3}\le M(t+1)^{-2/3},
\]
where the last inequality follows from convexity of $s\mapsto s^{-2/3}$. The triangle inequality in $L^2$ gives
\[
 \sqrt{\Exp[\norm{x^t-x^\star}^2]}\le\sqrt{a_t}+D\gamma_t\le(\sqrt M+D\gamma_0)t^{-1/3}.
\]
Thus $C_{\rm any}=(\sqrt M+D\gamma_0)^2$ works in~\eqref{eq:anytime-bound}, and~\eqref{eq:symmetric-exploitability-lipschitz} gives the exploitability bound.

To prove almost-sure convergence, fix $T\ge2$ and $\varepsilon>0$,
and let $\tau=\inf\{t\ge T:v_t\ge\varepsilon\}$, with
$\inf\varnothing=\infty$. For $N\ge T$, the event $\{\tau>t\}$
is $\mathcal F_t$-measurable, so~\eqref{eq:anytime-recursion} gives
\[
 \Exp[v_{\min\{\tau,N\}}]
 =\Exp[v_T]+\sum_{t=T}^{N-1}
   \Exp[\indic{\tau>t}(v_{t+1}-v_t)]
 \le MT^{-2/3}+H\sum_{t=T}^{\infty}t^{-5/3}.
\]
On $\{\tau\le N\}$, the stopped value is at least $\varepsilon$.
Applying Markov's inequality and letting $N\to\infty$ gives
\[
 \Prob(\exists\,t\ge T:\ v_t\ge\varepsilon)
 \le \frac{MT^{-2/3}+H\sum_{t=T}^{\infty}t^{-5/3}}{\varepsilon}
 \longrightarrow 0\qquad(T\to\infty).
\]
Applying this to $\varepsilon=1/j$ for every integer $j\ge1$
gives $v_t\to0$ almost surely. Finally,
$\norm{x^t-y^t}\le D\gamma_t\to0$, so both $y^t$ and $x^t$
converge to $x^\star$ almost surely.
\end{proof}

\section{Self-Play External Regret}
\label{app:regret}
\begin{proof}[Proof of Corollary~\ref{cor:symmetric-self-play-regret}]
We first bound the external pseudo-regret, defined by
\begin{equation}
 R_i(T)=\sum_{t=1}^T c_i(x_i^t,x_{-i}^t)-\min_{y_i\in\mathcal X}\sum_{t=1}^T c_i(y_i,x_{-i}^t).
 \label{eq:regret}
\end{equation}
For every sequence,
\[
\begin{aligned}
R_i(T)
&\le
\sum_{t=1}^T
\left[
c_i(x_i^t,x_{-i}^t)
-
\min_{y_i\in\mathcal X}
c_i(y_i,x_{-i}^t)
\right]\\
&=
\sum_{t=1}^T
\operatorname{gap}_i(x^t).
\end{aligned}
\]
For a lower bound, use the fixed comparator $x_i^\star$ in~\eqref{eq:regret}. Since $\norm{g_i(x^t)}\le\sqrt m\,\cmax$,
\[
 R_i(T)\ge\sum_{t=1}^T\ip{g_i(x^t)}{x_i^t-x_i^\star}\ge-\sqrt m\,\cmax\sum_{t=1}^T\norm{x^t-x^\star}.
\]
Together with the preceding upper bound and~\eqref{eq:symmetric-exploitability-lipschitz}, this gives $|R_i(T)|\le C_{\rm expl}\sum_{t=1}^T\norm{x^t-x^\star}$. By~\eqref{eq:anytime-bound} and Cauchy--Schwarz, its expectation is $O(\sum_{t=1}^T t^{-1/3})=O(T^{2/3})$.

For realized regret, write
\[
 A_T=\sum_{t=1}^T\bigl(\ip{Z_i^t}{d_i^t}-\ip{x_i^t}{g_i(x^t)}\bigr),\qquad N_T=\sum_{t=1}^T(d_i^t-g_i(x^t)).
\]
The increments of $A_T$ and each coordinate of $N_T$ have conditional mean zero by independence across players. Their absolute values are bounded by $m\cmax$ and $\cmax$, respectively. Orthogonality of martingale differences gives $\Exp|A_T|\le m\cmax\sqrt T$ and $\Exp\|N_T\|_1\le m\cmax\sqrt T$. Since $\mathcal X\subseteq[0,1]^m$, changing the cumulative cost vector by $N_T$ changes its minimum over fixed actions by at most $\|N_T\|_1$. Thus $|R_i^{\rm real}(T)-R_i(T)|\le|A_T|+\|N_T\|_1$, which, together with the pseudo-regret bound, proves the corollary.
\end{proof}

\section{A Fixed-Opponent Counterexample}
\label{app:counterexample}
\begin{proof}[Proof of Proposition~\ref{prop:symmetric-adversarial-counterexample}]
Consider two players and two links \(A,B\), with
\[
c_A(\ell)=c_B(\ell)=\ell.
\]
Suppose player \(2\) always selects link \(A\). Let
\[
x_1=(p,1-p),
\qquad
x_2=(1,0),
\]
where \(p\in[0,1]\) is player \(1\)'s probability of selecting \(A\).

Against this fixed opponent, player \(1\)'s conditional resource costs are
\[
g_A=2,
\qquad
g_B=1.
\]
Thus her unique best response in the original congestion game is to select
\(B\), i.e. \(p=0\).

The population average is
\[
m(x)
=
\frac{x_1+x_2}{2}
=
\left(
\frac{p+1}{2},
\frac{1-p}{2}
\right),
\]
and therefore
\[
x_1-m(x)
=
\left(
\frac{p-1}{2},
\frac{1-p}{2}
\right).
\]
Hence the difference between the \(A\)- and \(B\)-coordinates of the regularized gradient is
\[
\begin{aligned}
h(p)
&:=
(g_A-g_B)
+
\rho
\left[
\frac{p-1}{2}
-
\frac{1-p}{2}
\right]
\\
&=
1+\rho(p-1).
\end{aligned}
\]
For every \(\rho>1\), this field has the unique interior zero
\[
p^\rho
=
1-\frac1\rho
\in(0,1).
\]
Equivalently,
\[
h(p)
=
\rho(p-p^\rho),
\]
and therefore
\[
(p-p^\rho)h(p)
=
\rho(p-p^\rho)^2.
\]

Let
\[
\widehat h_t
=
\widehat G_{1,A}^t-\widehat G_{1,B}^t
\]
denote the corresponding semi-bandit estimate of the corrected
coordinate difference. By conditional unbiasedness of the estimator,
\[
\Exp[
\widehat h_t
\mid
\mathcal F_t
]
=
h(p_t).
\]
Write $r_t=y_{1,A}^t$ for the hidden probability and $p_t=(1-\gamma_t)r_t+\gamma_t q_{1,A}$ for the played probability. Euclidean projection onto the two-coordinate simplex gives
\[
 r_{t+1}=\Pi_{[0,1]}(r_t-\eta_t\widehat h_t/2).
\]
The single-player second-moment calculation in Lemma~\ref{lem:symmetric-consensus-estimator} gives $\Exp[\widehat h_t^2\mid\mathcal F_t]\le C/\gamma_t$ for some finite $C$, since $\widehat h_t^2\le2\norm{\widehat G_1^t}^2$. The calculation uses only player~$1$'s exploration floor, so it applies to the fixed opponent.

Let $a_t=\Exp[(r_t-p^\rho)^2]$ and $b_t=C\eta_t^2/(4\gamma_t)$. Expanding the squared distance and using $h(p_t)=\rho(r_t-p^\rho)+\rho(p_t-r_t)$, $|p_t-r_t|\le\gamma_t$, and $2ab\le a^2+b^2$ gives
\[
 a_{t+1}\le(1-s_t)a_t+s_t\gamma_t^2+b_t,\qquad s_t=\rho\eta_t/2.
\]
For all sufficiently large $N$, $0<s_t\le1$ for $t\ge N$. Iterating then gives
\[
 a_{t+1}\le a_N\prod_{j=N}^t(1-s_j)+\sup_{j\ge N}\gamma_j^2+\sum_{j\ge N}b_j .
\]
The product tends to zero because $\sum_j s_j=\infty$. Letting $N\to\infty$ proves $a_t\to0$. Since $|p_t-r_t|\le\gamma_t\to0$, also $\Exp[p_t]\to p^\rho$.

Against the fixed strategy of player $2$, choosing link $A$ incurs cost $2$, while choosing link $B$ incurs cost $1$. Hence $B$ is the best fixed action in hindsight, and $R_1^{\rm real}(T)$ equals the number of rounds in which player $1$ chooses $A$. Consequently,
\[
 \Exp[R_1^{\rm real}(T)]=\sum_{t=1}^T\Exp[p_t].
\]
Since $\Exp[p_t]\to p^\rho$, Ces\`aro convergence gives
\[
 \frac{\Exp[R_1^{\rm real}(T)]}{T}\longrightarrow p^\rho=1-\frac1\rho>0.
\]
\end{proof}
\end{document}